\documentstyle[twocolumn,eqsecnum,aps,epsfig]{revtex}
\begin{document}

\title{ Deuterons and space-momentum correlations in high energy nuclear collisions }

\author{B. Monreal$^{(a),\dag}$, S.A. Bass$^{(b),\ddag}$,
M. Bleicher$^{(c)}$, S. Esumi$^{(d)}$, W. Greiner$^{(c)}$,
Q. Li$^{(a)}$, H. Liu$^{(e)}$, W.J. Llope$^{(f)}$,
R. Mattiello$^{(g)}$, S. Panitkin$^{(e)}$, I. Sakrejda$^{(a)}$,
R. Snellings$^{(a)}$, H. Sorge$^{(h)}$, C. Spieles$^{(a)}$,
H. St\"ocker$^{(c)}$, J. Thomas$^{(a)}$, S. Voloshin$^{(a),\ast}$,
F. Wang$^{(a)}$, and N. Xu$^{(a)}$}

\address{\em (a) Nuclear Science Division, LBNL, Berkeley, CA 94720,USA} 
\address{\em (b) Department of Physics, Duke University, Durham, NC, 27708, USA}
\address{\em (c) Institute for Theoretical Physics, J.W. Goethe-University, D-60054, Frankfurt, Germany} 
\address{\em (d) Physics Institute, Heidelberg University, Philosophenweg 12, D-69120 Heidelberg, Germany}
\address{\em (e) Department of Physics, Kent State University, OH 44242, USA} 
\address{\em (f) T.W. Bonner Nuclear Laboratory, Rice University, TX 77005, USA} 
\address{\em (g) Niels Bohr Institute, Blegdamsvej 17, University of Copenhagen, DK-2100 Copenhagen, Denmark}
\address{\em (h) Department of Physics, SUNY at Stony Brook, NY 11794, USA} 

\date{\today}
\maketitle
\begin{abstract}
 Using a microscopic transport model together with a coalescence
 after-burner, we study the formation of deuterons in Au + Au central
 collisions at $\sqrt{s}=200$ $AGeV$. It is found that the deuteron
 transverse momentum distributions are strongly affected by the
 nucleon space-momentum correlations, at the moment of freeze-out,
 which are mostly determined by the number of rescatterings. This
 feature is useful for studying collision dynamics at ultrarelativistic
 energies.
\end{abstract}
%
%
\narrowtext

 Relativistic heavy ion collisions offer the unique opportunity to
 study hot and dense matter under controlled laboratory conditions
 (for recent reviews, see \cite{qm97,harris96a,bass98a,dumitru99}).
 However, the particle momentum distributions do not directly reveal
 the properties of the initial dense state of the system created in
 these collisions. The system undergoes longitudinal and transverse
 expansion which greatly affects the particle momentum
 distributions. Therefore the observation that for a given colliding
 system and within the same kinematic region, the slope
 parameter\footnote{The slope parameter is extracted from the fitting
 of the transverse mass distribution $\frac{1}{m_T}\frac{dN}{dm_T}$
 with an exponential function $A\cdot exp(-m_T/T)$. $A$ is a
 normalization parameter.} $T$ depends on the particle mass
 \cite{na44slope97} is of great interest and can be exploited to
 extract information about the reaction dynamics prior to the
 freeze-out stage. It has been found that the higher the particle
 mass, the larger the slope parameter is. This mass dependence is the
 strongest for the heaviest systems (Pb + Pb), and vanishes altogether
 in p+p collisions at similar energies \cite{na44slope97}.

 A related observation is that the size parameters $R_{T}$ and
 $R_{Long}$ ($R_{Long}$ and $R_T$ are the size parameters in the beam
 direction and perpendicular to the beam direction, respectively),
 which are extracted from two-particle correlation measurements,
 depend on the transverse momentum of the pair.  The higher the
 momentum, the smaller the size parameter
 \cite{na44size96,na49size96}. This dependence, like the slope
 parameter dependence, is the strongest (in $R_{Long}$) in the
 heaviest systems, although it is also observed in elementary
 collisions\cite{na22size97}.

 In any heavy-ion collision, the space-time freeze-out distribution
 and its dependence on the particle momentum are determined by the
 underlying dynamics.  The above observations are usually interpreted
 by considering nuclear fluid dynamics (NFD) type collective flow,
 which clearly leads to space-momentum
 correlations\cite{yuri88,chapman95,heinz87}. The NFD is not the
 only model for the interpretation of such correlation.  Another
 example might be string fragmentation \cite{anderson70}.

 Further insight into transverse collective flow can be gained from
 studying the mean transverse momentum, $\langle p_T \rangle$, of
 different particles like pions, kaons, and protons: specifically,
 analyzing the mean $p_T$ variation with particle mass.
 Unfortunately, the full space-momentum structure of the collision
 cannot be extracted from single-particle momentum spectra alone
 \cite{uli88,bass99}.  To shed more light on this issue, we propose to
 utilize deuteron distributions to investigate nucleon freeze-out
 properties.  In this letter we use the deuteron transverse momentum
 distribution as well as the ratio of the proton distribution to that
 of the deuteron \cite{uheinz99}, to extract information on
 space-momentum correlations and flow at RHIC energies.  In this
 study, we focus on central (impact parameter $b \leq$ 3.0 fm) Au + Au
 collisions at $\sqrt{s}=200$ $AGeV$.

 For our investigation we employ the Relativistic Quantum Molecular
 Dynamics model (RQMD)\cite{sorgehome,sorge95}. The model is well
 established and has been used successfully to describe many
 observables measured at AGS and SPS bombarding energies over a wide
 range of projectile-target combinations.  RQMD\cite{sorgehome} is a
 semi-classical microscopic approach which combines classical
 propagation with stochastic interactions.  Color strings and hadronic
 resonances can be excited in elementary collisions. Their
 fragmentation and decay lead to production of particles. Overlapping
 strings do not fragment independently from each other but form
 `ropes', chromoelectric flux-tubes whose sources are charge states in
 higher dimensional representations of color SU(3). RQMD is a full
 transport theoretical approach to reactions between nuclei (and
 elementary hadrons) starting from the initial state before overlap to
 the final state after the strong interactions have ceased
 (freeze-out). The model does not include light cluster
 productions. Therefore an after-burner, described in
 \cite{coal,negale94,sorgeplb95,mattiello95}, is used for deuteron
 yields calculations. More details of the coalescence type
 calculations can be found in \cite{negale94,mattiello95,mattiello97}
 and references therein.

 The deuteron binding energy ($\sim 2.2$ MeV)\cite{pdg98} is small
 compared to the characteristic freeze-out temperature ($\sim 140$
 MeV) of ultra-relativistic heavy-ion collisions we are studying.
 Hence, deuterons cannot survive rescattering.  Since many
 rescatterings occur within the hot and dense reaction phase, the only
 deuterons that survive and escape are those formed near the
 freeze-out stage, either on the surface of the fireball or at a later
 time when the environment is dilute.

 The particle phase-space distribution at freeze-out reflects physics at
 earlier stage of the collision. Similar to the two-particle
 correlation functions, the probabilities of bound state (deuterons
 and heavier clusters) formation are determined by this distribution
 \cite{lednicky82}. 

 The sensitivity of the two particle correlation measurement to the
 source size decreases when the source size (and/or duration of the
 emission) becomes large. The analysis of the deuteron yield relative
 to the proton yield has no such loss in sensitivity although the
 value of the deuteron yield decreases due to larger distance between
 the two nucleons.


 Assuming a Gaussian form for the nucleon
 source \cite{mrowczynski93,leupold94}, a size parameter $R_g$ can be
 extracted from the single particle distributions of protons and
 deuterons \cite{llope95}:

\begin{equation}
R_g^3 = \frac{3}{4}(\sqrt{\pi}\hbar c)^3 \frac{m_d}{m_p^2}\frac{(E_p
\frac{d^3N_p}{d^3p})^2}{E_d \frac{d^3N_d}{d^3p}}, 
\end{equation}

\noindent{where $m_p, m_d$ are the proton and the deuteron masses,
 respectively.  The invariant distribution is $E_id^3N_i/d^3p$ with
 ($i=proton,deuteron)$.  The above equation assumes that the deuteron
 energy is the sum of the proton and the neutron energies and that
 there is no space-momentum correlation in particle distributions at
 freeze-out.  Then the space-momentum correlation can be studied by
 inspecting the $R_g$ as a function of the transverse momentum $p_T$.}

 The Gaussian size parameter $R_g$ as a function of the nucleon
 transverse mass $m_T$ is shown in figure 1.  Here, the filled circles
 represent the results from original(default) RQMD events. The open
 symbols represent events with altered space-momentum correlation. The
 squares represent the so called aligned case, where for each nucleon
 the space vector $\vec{r_T}$ is aligned with the transverse momentum
 vector $\vec{p_T}$. The triangles represent the case where the angle
 between $\vec{p_T}$ and $\vec{r_T}$ has been randomized. Note that in
 the aligned and random cases only the relative orientation of
 $\vec{r_T}$ to $\vec{p_T}$ is modified: momentum distributions and
 projections onto either ${r_T}$ or ${p_T}$ are not touched. In the
 randomized case, the amplitudes of vectors $|\vec{r_T}|$ and
 $|\vec{p_T}|$ are still correlated. To remove such correlation, the
 vectors $\vec{r_T}$ and $\vec{p_T}$ were scrambled (open circles in
 Fig.1). After the operation, the correlations among $\vec{r_T}$ and
 $\vec{p_T}$ are removed and, as expected, the distribution is almost
 flat as a function of $m_T$. To guide the eye, the solid line
 represents the function $9.75 \cdot (m_T)^{1/2}$.

 While no large differences are observed between the normal and the
 aligned results, a dramatic effect is evident between the randomized
 and normal cases. This implies that nucleon momenta are already
 largely aligned in the real events. The calculated mean cosine of the
 angle between transverse space and momentum vector is about $\langle
 {\rm cos}(\theta) \rangle \approx$ 0.9 at mid-rapidity.  As one can
 see in the figure, all distributions converge to a point where $R_g
 \approx 9.5$ (fm) at $m_T \rightarrow m_p$ ($p_T \rightarrow 0$),
 indicating that the `true source size' can be measured at small
 $p_T$, while at higher $p_T$ the size parameter is found to be
 sensitive to space-momentum correlations.  As it was mentioned above,
 the deuterons are sensitive to the space-time-momentum correlations
 in the same fashion as the two particle correlation function. Indeed,
 these effects were also found in the study of two pion correlation
 functions\cite{esumi97,fields95}.


 Randomizing (or aligning) the nucleon space and momentum vectors
 changes the space-momentum correlations. It affects both the deuteron
 yields and momentum distributions~\cite{mattiello95}. This effect is
 vividly displayed in figure 2, where proton (circles) and deuteron
 (squares) average transverse momenta $\langle p_T \rangle$ are shown
 as a function of rapidity. Plots (a), (b), and (c) are for normal,
 randomized, and aligned cases, respectively; plot (d) is the results
 from a calculation without rescatterings among baryons (rescattering
 here means interaction with produced particles). The solid- and
 dash-lines in (a) and (d) represent the values of $\langle p_T
 \rangle$ for kaons and pions, respectively.

 At mid-rapidity the values of the average transverse momentum
 $\langle p_T \rangle$ of pions, kaons, and protons are $\langle p_T
 \rangle$= 0.4, 0.6, and 0.85 GeV/c, respectively. The value for
 deuterons is about 1.4 GeV/c, see Fig. 2(a). These values are very
 similar to the results in References of \cite{dumitru99,bass99}.  The
 difference in mean $p_T$ between proton and deuteron decreases from
 about 550 to 150 MeV/c as one moves away from mid-rapidity to $y \geq
 4.$ After randomization, Fig. 2(b), the splitting between deuterons
 and protons in the mean transverse momentum becomes constant $\sim
 150$ MeV/c. For the aligned case, the difference changes from 580 to
 280 MeV/c from mid-rapidity to $y \geq 4.$ The change in transverse 
 momentum at mid-rapidity is about 30 MeV/c, but at $y \geq 4$, the
 difference is about 100 MeV/c compared to the normal case. Similar to
 the random case, the results of calculations without baryon
 rescattering (Fig. 2(d)) show a constant difference between deuteron
 and proton transverse momentum of about 150 MeV.


 Given the distributions shown in Fig. 1 and Fig. 2, one may discuss
 the physics in terms of collectivity. To proceed, we evaluate the
 average transverse velocities of pions, kaons, and nucleons. In the
 following analysis, the collective velocity is defined as:

\begin{equation}
\langle \beta_T \rangle =\langle  \frac{\vec{p_T}\cdot\vec{r_T}}{r_T m_T} \rangle.
\end{equation}

 Figure 3 (a)-(d) show the velocities as a function of rapidity for
 different cases and Fig, 3(e) and (f) depict the mean number
 of nucleon collisions as a function of rapidity.

 Fig. 3 (a) shows that on average particles of different types are
 moving together with a similar velocity. These results indicate a
 certain amount of collectivity in Au + Au central collisions at RHIC
 energies.\footnote{The same model calculations, for Au + Au collisions
 at lower energies (few GeV per nucleon), indicate no such collective
 behavior although protons and light nuclear cluster distributions do
 show characteristics of collective motion at SIS and Bevalac energies
 (see \cite{hgritter97} and references therein).}  On the other hand,
 the RQMD calculations predict the charged pion ($\pi ^+$ + $\pi ^-$)
 to nucleon ($p$ + $n$) ratio is about 10 near mid-rapidity for the
 central Au + Au collisions at the RHIC energy. L\'evai and M\"uller
 argue that in such baryon-poor region the equal magnitude of pion
 and nucleon flow velocities can be established only at an earlier
 deconfined phase \cite{levai91}. Such deconfined phase, of course,
 was not included in the present calculation. The other interesting
 observation in Fig. 3 is that at a given rapidity, the values of the
 velocity are closely correlated with the number of rescatterings: the
 larger the number of collisions, the higher the velocity (see
 Fig. 3(a) and (e)). In the case of no-rescattering calculation (see
 Fig. 3(f)), the number of collisions is about 2.5 and the
 velocity is about zero over the whole rapidity range
 (Fig. 3(d)). Recall, that, in the no-rescattering case, the splitting
 in the average transverse momentum  according to mass vanishes,
 see Fig. 2(d). Once the space-momentum correlation is altered, 
 the collectivity is destroyed (Fig. 3(c)).

 Note that RQMD predicts the averaged transverse collective velocity
 of mid-rapidity about 0.6 at the RHIC energy (Fig. 3(a)) while at the
 SPS ($\sqrt{s} \approx 20$ $AGeV$) the value is about 0.4-0.45$c$.


 In summary, using a microscopic transport model RQMD(v2.4) and a
 coalescence after-burner, we have studied the transverse momentum
 distributions of pions, kaons, nucleons, and deuterons in different
 rapidity regions for central Au + Au ion collisions at $\sqrt{s} =
 200$ $AGeV$.  Employing the deuteron as a probe, we have demonstrated
 that a large number of rescatterings leads to the space-momentum
 correlation at freeze-out and is responsible for the decrease of the
 ratio of $N^2(proton)/N(deuteron)$ as a function of $m_T$.  Should
 new physics occur at RHIC energy, a modification of the
 space-momentum structure will manifest itself in the deuteron yields
 and its transverse momentum distributions. These distributions can be
 measured in the STAR TPC and other RHIC experiments.

 $\dag$ B. Monreal, from Yale University, is at the Lawrence
 Berkeley National Laboratory through the Center for Science and
 Engineering Education.

 $\ddag$ Feodor Lynen Fellow of the A.v. Humboldt Foundation.

 $\ast$ On leave from Moscow Engineering Physics Institute, Moscow,
115409, Russia.
 
 We are grateful for many enlightening discussions with
 Dr. A. Dumitru, Dr. U. Heinz, Dr. D. Keane, Dr. S. Pratt, and
 Dr. H.G. Ritter.  We thank Dr. J. Nagle for the use of the
 coalescence code. This research used resources of the National Energy
 Research Scientific Computing Center.  This work has been supported
 by the U.S. Department of Energy under Contract No. DE-AC03-76SF00098
 and W-7405-ENG-36 and the Energy Research Undergraduate Laboratory
 Fellowship and National Science Foundation.


\begin{center}

\begin{figure}[hct]
\centerline{\epsfig{figure=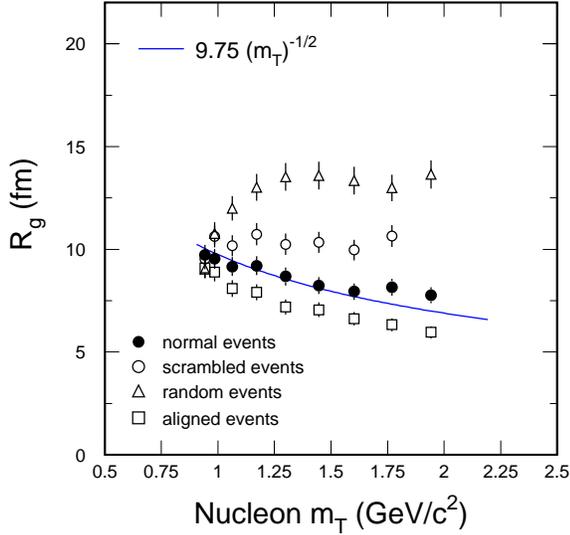,width=9.5cm}}
\vspace{-0.5cm}
\caption{Gaussian radius parameter $R_g$ as a function of proton
transverse mass $m_T$ for Au + Au central collisions ($b \leq 3 $fm)
at $\sqrt{s}=200$ $AGeV$. Filled circles represent the results for
original RQMD events. Open symbols represent the results for the
events where the correlation between $\vec{p_T}$ and $\vec{r_T}$ has
been altered.  Only mid-rapidity ($|y|\leq 1.0$) nucleons are used for
the plot. }
\end{figure}

\begin{figure}[h]
\centerline{\epsfig{figure=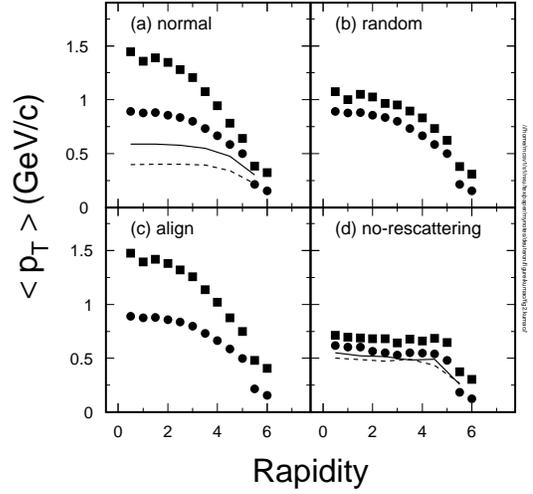,width=8.6cm}}
\caption{ Deuteron (filled square), nucleon (filled circle), kaon
(solid-line), and pion (dashed-line) mean transverse momentum as a
function of rapidity.}
\end{figure}

\vspace{-0.5cm}
\begin{figure}[h]
\centerline{\epsfig{figure=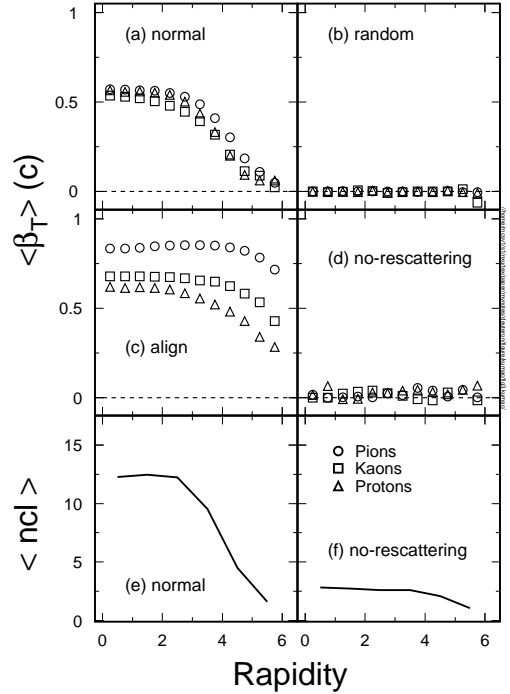,width=8.5cm}}
\caption{ RQMD mean velocities $\langle \beta_T \rangle$ of pions,
kaons and nucleons and mean number of collisions $\langle ncl \rangle$
for nucleons as a function of rapidity. }
\end{figure}

\end{center}          
\end{document}